# Meshless simulation for thermo-mechanical properties of single-walled carbon nanotubes based on the thermal-related higher order Cauchy-Born rule

Xu Guo*, Xiangyang Wang

*State Key Laboratory of Structural Analysis for Industrial Equipment*

*Department of Engineering Mechanics*

*Dalian University of Technology, Dalian 116024, China*

**Abstract**

In the present paper, a temperature-dependent meshless numerical framework based on the thermo-related quasi-continuum constitutive model is developed for predicting the thermal mechanical properties of single-walled carbon nanotubes (SWCNTs) at finite temperature. The extended thermal-related higher order Cauchy-Born (THCB) rule included second order deformation gradient relates the deformation of bond vectors of the atomic system and that of the continuous medium, which can capture the curvature effect of carbon nanotubes (CNTs) conveniently. Helmholtz free energy is employed to allow for the thermal effect of SWCNTs. In the meshless numerical implementations of the theory, the Newton iteration method is applied to find the equilibrium configuration of a SWCNT subjected to large deformation at a prescribed temperature only with the nodal displace parameters as optimization variables. The finite deformation behaviors of armchair and zigzag SWCNTs under axial compression and torsion are tested. It is shown that the simulation results are in good agreement with those obtained by molecular dynamic methods even with fewer meshless nodes used.

* Corresponding author. Tel: +86-411-84707807. *E-mail addresses:* guoxu@dlut.edu.cn

## 1. Introduction

Nowadays, as a new allotrope of carbon, carbon nanotubes (CNTs) have been attracting intense research partly due to their exceptional mechanical and electronic properties. Many applications of CNTs have been reported such as nano-electronics, quantum wires, composites, biosensors, etc. In order to make good use of these nanoscale materials, it is imperative to have a good knowledge of their thermal mechanical responses.

In-situ experiments are necessary means to obtain the intrinsic mechanical properties of CNTs (Treacy et al., 1996; Krishnan et al., 1998; Wong et al., 1997; Yu et al., 2000). However, it is a great challenge to manipulate such small objects (with nanometer scale) properly with atomic force microscopy or transmission electron microscopy. And the experimental results are highly dependent on samples such as with or without defects. In addition, atomistic modeling approaches (Yakobson et al., 1996; Iijima et al., 1996; Popov et al., 2000; Hernández et al., 1998; Goze et al., 1999; Liu et al., 2004) such as molecular dynamics methods are widely used for predicting the mechanical properties of CNTs, which can capture detailed knowledge of atoms of the considered atomic system such as positions and velocities. However, the efforts of computation are tremendous especially the considered system is relatively larger, which is a fatal barrier from the engineering application point of view. Fortunately, the continuum-mechanical theory can be used as an appropriate candidate for predicting the mechanical properties of CNTs, which has been proved to be of efficiency and reliability recently. Govindjee and Sackman (1999) studied the validity of the simple elastic sheet model for investigating the mechanical properties of nanotubes. It is shown that the analysis results of mechanical response for nanotubes subjected to bending loads were dependent on the considered system size. Ru (2000a, b) pointed out that the effective bending stiffness of SWCNTs should be regarded as an independent material parameter in the developed elastic honeycomb model for analyzing the mechanical properties of nanostructures. Furthermore with the same theory, they investigated the elastic buckling behavior of single-walled carbon nanotube ropes. Odega et al. (2002) applied the proposed equivalent-continuum model to predict the effective-continuum geometry of nano-sized objects and bending rigidity of graphene sheet. Subsequently, it is found by Sears and Batra (2004, 2006) that the strain energy of SWCNTs under bending deformations based on the Euler-Bernoulli beam theory were in good agreement with the results obtained by the molecular-mechanics simulations and the buckling patterns of multi-walled carbon nanotubes (MWCNTs) with large aspect ratio were comparable to those simulated through the Euler buckling theory. In addition, they also established a finite element numerical framework for simulating the buckling behavior of MWCNTs. It should be pointed out that the continuum mechanical models mentioned above are phenomenological methods and a few additional material parameters need to be obtained by fitting of numerous experimental data which sometimes are unavailable.

Recently, a quasi-continuum modeling approach (Tadmor et al., 1996, 1999; Gao and Klein, 1998; Miller et al., 1998; Shenoy et al., 1999; Zhang et al., 2002a, b) for simulating mechanical respones of nanometer-sized materials has been used widely. It can be extracted directly from atomic crystalline system based on the empirical potential with the Cauchy-Born rule relating the deformation of the atomic configuration and that of continuum. Compared with the conventional continuum models, the quasi-continuum based model does not need any fitting parameters inputting. Arroyo and Belytschko (2002, 2004) first pointed out the lattice vectors being on chords of a curved crystalline film rather than lying on its tangent space. Accordingly, the classical Cauchy-Born rule can't be employed directly for CNTs because it cannot take the curvature effect into consideration. Under this circumstance, they developed a finite deformation quasi-continuum model based on the extended exponential Cauchy-Born rule which naturally maps a tangent vector of a manifold into itself. In addition, they also established a numerical simulation framework based on their proposed constitutive model for simulating the finite deformation behaviors of CNTs without considering the temperature effect. Sunyk and Steinmann (2003) proposed a continuum-atomistic model for investigating the inhomogeneous deformation response of atomic crystalline. Based on the Cauchy-Born rule with second-order deformation gradient incorporating, this model was employed to investigate non-homogeneous simple shear deformation of the considered atomic film. Guo et al. (2006a, b) developed a nanoscale quasi-continuum consititutive model for simulating the mechanical properties of SWCNTs based on the so-called higher order Cauchy-Born (HCB) rule and Tersoff-Brenner interatomic potential. Due to including the second-order deformation gradient tensor in HCB rule, the positions of carbon atoms on a deformed crystalline membrane can be found with more accuracy, at the same time, the curvature effect of CNTs can be considered in a convenient way. Sun and Liew (2008a, b) proposed a HCB rule based mesh-free computational framework to analysis the finite deformation behaviors of SWCNTs. As expected, their simulation results for boundary problems of SWCNTs under large deformations were in consistent with those obtained by atomistic model. However, it is worth noting that none of these quasi-continuum models proposed above allowed for the temperature effect.

It is well-known that, cooperating with other facilities or independence, the real nano-structures or devices will work in a temperature environment. Therefore, an effective and convenient computational model for predicting the thermal mechanical properties of CNTs should be established by taking account of the temperature effect. To the best of our knowledge, there are few efforts on this aspect which can be found in the literature. A finite-temperature quasi-continuum constitutive model was developed by Jiang et al. (2004, 2005) for predicting the thermal mechanical properties of SWCNTs at finite temperature. This model was derived from interatomic potential with the standard Cauchy-Born rule as the linkage. The Helmholtz free energy was used as thermodynamic potential with incorporating the local harmonic approximation. As applications of the model, they investigated several temperature-dependent mechanical properties of graphene sheet and SWCNTs, such as coefficients of thermal expansion, specific heat and Young's modulus. The results attest to the efficiency of their proposed model.

In this study, a temperature-dependent mesh-free implementation scheme based on the thermo-related constitutive model, with THCB rule being the key component, is proposed for investigating the thermal mechanical properties of SWCNTs under large deformations at finite temperature. The equivalent-continuum surface free energy density is obtained through the homogenization of the free energy over the representative cell in a virtual graphene sheet which is formed by the positions of vibration centers of carbon atoms. Because of introducing the higher order deformation gradient in the kinematic description, from the numerical implementation point of view, the approximation displacement function over the domain of the considered system requires at least $C^1$-continuity. Fortunately, the mesh-free method (Belytschko et al. 1996, 1998) can be employed for the hyperelastic constitutive model since it possesses non-local properties (Sun and Liew, 2008a, b). In this study, the Newton iteration method is used to find the equilibrium configuration in each loading step. As the unique optimization variables, the nodal displacements can be obtained rapidly through several iterations for the discrete nonlinear stiffness equations. Finally, as applications of the developed thermal-mechanical numerical framework, we study the buckling and post-buckling behaviors of SWCNTs which are subjected to compression and torsion deformations at different temperatures. The numerical simulation results show that, with fewer meshfree nodes used, the deformation modes and energy of SWCNTs match well with those obtained by molecular dynamic methods. Furthermore, the variations of critical strain of zigzag and armchair SWCNTs with different radii at 0K (non-thermal effect) and 300K (room temperature) are also investigated. These numerical simulation results demonstrate the feasibility of our developed thermal-mechanical quasi-continuum model.

The rest of the paper is organized as follows: The Tersoff-Brenner interatomic potential is firstly introduced in Section 2. Section 3 presents the proposed THCB rule. Section 4 describes the Helmholtz free energy and introduces the deformation mapping relationship from an undeformed virtual graphene sheet to the corresponding deformed hollow cylindrical SWCNT. In Section 5, we deduce the temperature-related constitutive model based on THCB rule. Subsequently, the meshless numerical simulation framework is proposed in Section 6. Based on the above preparations, several numerical examples for thermal-mechanical properties of SWCNTs are presented in Section 7. In the end, some concluding remarks are given in Section 8.

## 2. Tersoff-Brenner interatomic potential for carbon

The multi-body interatomic potential which was proposed by Tersoff (1988) and Brenner (1990) is widely used for the description of the interactions between carbon atoms. It has the following form:

$$V(r_{IJ}) = V_R(r_{IJ}) - \bar{B}_{IJ} V_A(r_{IJ}) \tag{1}$$

where $r_{IJ}$ denotes the distance between atoms $I$ and $J$; $V_R$ and $V_A$ represent the repulsive and attractive pair potentials between bonded carbon atoms; $\bar{B}_{IJ}$ is the multi-body potential depending on the bond length $r_{IJ}$ as well as the bond angles between $IJth$ bond vector and other adjacent bonds emanating from atom $I$.

$$V_R(r) = \frac{D^{(e)}}{S-1} e^{-\sqrt{2S}\beta[r-R^{(e)}]} f_c(r) \tag{2}$$

$$V_A(r) = \frac{D^{(e)}}{S-1} S e^{-\sqrt{2/S}\beta[r-R^{(e)}]} f_c(r) \tag{3}$$

$$\bar{B}_{IJ} = \frac{1}{2}(B_{IJ} + B_{JI}) \tag{4}$$

$$B_{IJ} = \left[1 + \sum_{K(\neq I,J)} G(\theta_{IJK}) f_c(r_{IK})\right]^{-\delta} \tag{5}$$

$$G(\theta) = a_0 \left[1 + \frac{c_0^2}{d_0^2} - \frac{c_0^2}{d_0^2 + (1 + cos\theta)^2}\right] \tag{6}$$

with $f_c$ denoting the smooth switch function to limit the range of the potential. It is expressed as follows:

$$f_c(r) = \begin{cases} 1 & r \le R^{(1)} \\ \frac{1}{2}\left\{1 + \cos\left(\frac{\pi\left(r - R^{(1)}\right)}{R^{(2)} - R^{(1)}}\right)\right\} & R^{(1)} < r \le R^{(2)} \\ 0 & r > R^{(2)} \end{cases} \tag{7}$$

the parameters $D^{(e)}$, $S$, $\beta$, $R^{(e)}$, $\delta$, $a_0$, $c_0$, $d_0$, $R^{(1)}$ and $R^{(2)}$ have been determined by Brenner (1990) though fitting the binding energy and lattice constants of carbon allotropes. In this paper, the parameters employed in the following sections are given as

$D^{(e)} = 6.000eV,\ S = 1.22,\ \beta = 21nm^{-1},\ R^{(e)} = 0.1390nm,\ \delta = 0.50000,\ a_0 = 0.00020813,$ $c_0 = 330,\ d_0 = 3.5,\ R^{(1)} = 0.17nm,\ R^{(2)} = 0.20nm.$

## 3. Thermal-related higher order Cauchy-Born rule for thermal mechanical properties of CNTs

### 3.1 Higher order Cauchy-Born rule for CNTs

Generally, the standard Cauchy-Born rule is applied to establish constitutive models for investigating the mechanical properties of nano-scaled carbon materials in the principle of quasi-continuum framework (Tadmor et al., 1996, 1999; Gao and Klein, 1998; Shenoy et al., 1999; Zhang et al., 2002a, b). According to its definition, the standard Cauchy-Born rule expresses the deformation of lattice vector in bulk materials as (see Fig. 1 for an illustration)

$$\boldsymbol{r} = \boldsymbol{F}_0 \cdot \boldsymbol{R} \tag{8}$$

where $\boldsymbol{R}$ is an undeformed lattice vector in the reference configuration and $\boldsymbol{r}$ is the corresponding vector in the current deformed configuration, $\boldsymbol{F}_0$ represents the two point deformation gradient tensor at absolute zero temperature. It is well-known that CNTs are curved crystalline membranes with one (SWCNTs) or several atomic thickness (MWCNTs). The classical Cauchy-Born rule, however, could only maps the lattice vectors in the tangent space of the curved crystalline film. In fact, the lattice bonds are on the chords of the curved manifold pointed by Arroyo and Belytschko (2002, 2004). If the quasi-continuum constitutive model is directly constructed based on the standard Cauchy-Born rule, the numerical results for CNTs will be unstable and unphysical (Sun and Liew 2008a, 2008b). From this point of view, a so-called higher order Cauchy-Born (HCB) rule proposed by Guo et al. (2006) for CNTs came into being

$$\boldsymbol{r} = \boldsymbol{x}(\boldsymbol{X}+\boldsymbol{R}) - \boldsymbol{x}(\boldsymbol{X}) \cong \boldsymbol{F}_0(\boldsymbol{X})\cdot\boldsymbol{R} + \frac{1}{2}\boldsymbol{G}_0(\boldsymbol{X}):(\boldsymbol{R}\otimes\boldsymbol{R}) \tag{9}$$

where $\boldsymbol{r}$ and $\boldsymbol{R}$ are the lattice vectors in the reference and current configuration, respectively. $\boldsymbol{X}$ is the material coordinate vector of a point in the undeformed configuration, and $\boldsymbol{x}$ is the corresponding coordinate vector in the current configuration. $\boldsymbol{F}_0(\boldsymbol{X})$ and $\boldsymbol{G}_0(\boldsymbol{X}) = \nabla\boldsymbol{F}_0$ are, respectively, the first and second order deformation gradient tensors which are functions of $\boldsymbol{X}$ at absolute zero temperature.

Additionally, due to the microstructure of graphene sheet isn't centro-symmetry, an inner relaxation parameter vector $\boldsymbol{\xi}$ (Tadmor et al., 1999) should be introduced in the reference configuration to ensure the internal equilibrium of the non-centrosymmetry atomic lattice structure (see Fig. 2 for reference). Therefore HCB rule for CNTs should be rewritten as

$$\boldsymbol{r} = \boldsymbol{F}_0\cdot(\boldsymbol{R}+\boldsymbol{\xi}) + \frac{1}{2}\boldsymbol{G}_0:[(\boldsymbol{R}+\boldsymbol{\xi})\otimes(\boldsymbol{R}+\boldsymbol{\xi})] \tag{10}$$

Based on the above HCB rule, a hyper elastic constitutive model can be established for analyzing the mechanical properties of CNTs. It is confirmed that, with including the second order deformation gradient term in kinematic description and taking the inner relaxation parameters into consideration, the accuracy of the simulations for predicting the mechanical properties of SWCNTs can be enhanced (Guo et al. 2006a, b) and, at the same time, less computational efforts are consumed (Sun and Liew 2008a, b).

3.2 Thermal-related higher order Cauchy-Born rule for CNTs

Obviously, carbon atoms move quickly around their vibration centers all the time in temperature environment. It is assumed that, if the temperature is not extremely high, it could only affect the variation of the bond length rather than change the crystalline lattice structure of CNTs. Based on this assumption, the virtual lattice structure is also of honeycomb shape which is formed by the vibration centers of atoms (see Fig. 3 for reference). Under this circumstance, the corresponding thermal-related higher order Cauchy-Born (THCB) rule can be proposed as

$$\boldsymbol{r}_{IJ}^{c}(T)=\boldsymbol{F}\cdot\left(\boldsymbol{R}_{IJ}^{c}(T)+\boldsymbol{\xi}_{I}^{c}(T)\right)+\frac{1}{2}\boldsymbol{G}:\left[\left(\boldsymbol{R}_{IJ}^{c}(T)+\boldsymbol{\xi}_{I}^{c}(T)\right)\otimes\left(\boldsymbol{R}_{IJ}^{c}(T)+\boldsymbol{\xi}_{I}^{c}(T)\right)\right] \tag{11}$$

where $\boldsymbol{R}_{IJ}^{c}(T)$ and $\boldsymbol{r}_{IJ}^{c}(T)$ are virtual bond vectors between the vibration centers of atoms $I$ and $J$ in the initial undeformed configuration and current deformed structure at prescribed absolute temperature $T$ respectively. And $\boldsymbol{\xi}_{I}^{c}(T)$ is the inner displacement vector between two virtual central symmetry sublattice structures of the atom $I$ at temperature $T$. $\boldsymbol{F}$ and $\boldsymbol{G}$ are the first and second order deformation gradient tensors in the temperature field. And the precise derivations of these deformation gradient tensors for SWCNTs are given in the next section.

## 4. Helmholtz free energy and the deformation mapping

### 4.1 Helmholtz free energy

It is worth noting that when analyzing the thermal mechanical properties of condensed materials such as CNTs, Helmholtz free energy is an appropriate choice. From the theory of statistical mechanics, the free energy $F_e$ of the considered system can be expressed as

$$F_e = -k_B T \ln Z \tag{12}$$

where $k_B = 1.38 \times 10^{-23} J \cdot K^{-1}$ is the Boltzmann constant; $T$ is the absolute temperature; and $Z$ denotes the partition function which can be written as

$$Z = \sum_r e^{-\beta E_r} \tag{13}$$

with $r$ expressing the $r-th$ energy state of the considered system; $\beta = 1/(k_B T)$; and $E_r$ is the total energy of the $r-th$ energy state. It is assumed herein that the vibration amplitudes of the atoms are not large if the prescribed temperature is not very high. Under this assumption, by including the quasi harmonic assumption (QHA), the total energy $E_r$ can be calculated through the following approximate form

$$E_r \cong \sum_{k=1}^{3N} \hbar\omega_k \left(n_k^r + \frac{1}{2}\right) + U_0\,, r = 0,1,2,\ldots \tag{14}$$

where $N$ is the total number of carbon atoms in the considered nano-scaled object. $\hbar$ is the Planck's constant $1.05 \times 10^{-34} J \cdot s$ . $\omega_k, (k = 1, \ldots, 3N)$ is the $k-th$ order vibration frequency. $n_k^r$ denotes the quantum number associated with the $r-th$ micro-state of the system for the $k-th$ particle. $U_0$ is the total interatomic potential of the considered system. From Eq. (12) to Eq. (14), the free energy of the system can be described precisely as

$$F_e = U_0 + k_B T \sum_{k=1}^{3N} \ln\left(2 \cdot \sinh\left(\frac{\hbar\omega_k}{2k_B T}\right)\right) \tag{15}$$

where the vibration frequency $\omega_k$ can be calculated by the following eigenvalue problem

$$\left| \frac{1}{m_c} \frac{\partial^2 U_0}{\partial \boldsymbol{x} \partial \boldsymbol{x}}\bigg|_{\boldsymbol{x}=\boldsymbol{x}^0} - \boldsymbol{I}_{3N\times 3N}\omega^2 \right| = 0 \tag{16}$$

where $\boldsymbol{x}^0$ denotes the position of vibration centers of all the atoms, $m_c$ is the carbon atom mass, and $\boldsymbol{I}_{3N\times 3N}$ denotes the $3N \times 3N$ identity matrix.

Obviously, if the considered atomic system is relatively large, to find all the vibration frequencies via solving the $3N \times 3N$ characteristic determinant in Eq. (16) is quite time-consuming. In order to overcome this difficulty, the so-called local harmonic approximation (LHA) is adopted to neglect the vibration coupling effect between different atoms. Under this circumstance, the order of the matrix in Eq. (16) will decrease to $3 \times 3$

$$\left| \frac{1}{m_c} \frac{\partial^2 U_0}{\partial \boldsymbol{x}_I \partial \boldsymbol{x}_I}\bigg|_{x_I=x_I^0} - \boldsymbol{I}_{3\times 3}\left(\omega_{I\beta}\right)^2 \right| = 0 \tag{17}$$

where $\omega_{I\beta}, (I = 1, \dots, N, \beta = 1,2,3)$ is the $\beta - th$ degree of freedom of the vibration frequency of atom $I$ with other adjacent atoms fixed, $\boldsymbol{I}_{3\times 3}$ denotes the $3 \times 3$ unit matrix.

From the above assumption, the free energy can be rewritten as

$$F_e = U_0 + k_B T \sum_{I=1}^{N} \sum_{\beta=1}^{3} \ln\left(2 \cdot \sinh\left(\frac{\hbar\omega_{I\beta}}{2k_B T}\right)\right) \tag{18}$$

As pointed out by Foiles (1994), with the LHA employed, the accuracy of the simulation results for analyzing the thermal mechanical properties of SWCNTs are acceptable at lower temperature (below one half of the melting point), due to the vibration anharmonic effect of atoms increasing with temperature.

4.2 The deformation mapping

Ordinarily, it is assumed that SWCNTs can be obtained by rolling up to seamless hollow cylinders by planar graphite sheets. It is known that along the different rolling directions, SWCNTs can be divided into three categories: zigzag, armchair and chiral nanotubes (Saito et al. 1992).

First of all, we select a planar virtual graphite sheet as the reference configuration (see Fig. 4 (a)). A longitudinal stretch parameter $\lambda_1(T)$ and a circumferential stretch parameter $\lambda_2(T)$ as well as a twisting angle per length $\theta(T)$ at absolute temperature $T$ should be introduced for the purpose of guaranteeing the inner equilibrium of the undeformed SWCNT. From the preparations made above, the deformation mapping relationship from Fig. 4 (a) to (b) can be derived exactly as follows

$$x_1 = \lambda_1(T)X_1 \tag{19a}$$

$$x_2 = \lambda_2(T)R(T)\sin\left(\frac{X_2}{R(T)} + \theta(T)\lambda_1(T)X_1\right) \tag{19b}$$

$$x_3 = \lambda_2(T)R(T)\left(1 - \cos\left(\frac{X_2}{R(T)} + \theta(T)\lambda_1(T)X_1\right)\right) \tag{19c}$$

where $X_1$ and $X_2$ are the material coordinates of a point in the reference configuration. And $x_1$, $x_2$ and $x_3$ are the corresponding coordinates of the undeformed SWCNT in the Euler coordinate system. $R(T)$ is the radius of the undeformed SWCNT at temperature $T$. It can be calculated by

$$R(T) = \frac{a_0(T)\sqrt{3(n^2 + nm + m^2)}}{2\pi} \tag{20}$$

where $n$ and $m$ are a pair of integer parameters which are employed to describe the chiral characterization of CNTs. $a_0(T)$ is the virtual bond length of the graphite sheet(see Fig. 3 for reference).

The first and second order deformation gradient tensors $\boldsymbol{F}$ and $\boldsymbol{G}$ in the case of temperature $T$ can be derived from the above parameterized deformation mapping relationship in Eq. (19)

$$F_{iJ} = \frac{\partial x_i}{\partial X_J} = \begin{bmatrix} \lambda_1(T) & 0 \\ \theta(T)\lambda_1(T)\lambda_2(T)R(T)A_2 & \lambda_2(T)A_2 \\ \theta(T)\lambda_1(T)\lambda_2(T)R(T)A_1 & \lambda_2(T)A_1 \end{bmatrix} \tag{21}$$

$$G_{iJ1} = \frac{\partial F_{iJ}}{\partial X_1} = \begin{bmatrix} 0 & 0 \\ -\theta^2(T)\lambda_1^2(T)\lambda_2(T)R(T)A_1 & -\theta(T)\lambda_1(T)\lambda_2(T)A_1 \\ \theta^2(T)\lambda_1^2(T)\lambda_2(T)R(T)A_2 & \theta(T)\lambda_1(T)\lambda_2(T)A_2 \end{bmatrix} \tag{22a}$$

$$G_{iJ2} = \frac{\partial F_{iJ}}{\partial X_2} = \begin{bmatrix} 0 & 0 \\ -\theta(T)\lambda_1(T)\lambda_2(T)A_1 & -\dfrac{\lambda_2(T)}{R(T)}A_1 \\ \theta(T)\lambda_1(T)\lambda_2(T)A_2 & \dfrac{\lambda_2(T)}{R(T)}A_2 \end{bmatrix} \tag{22b}$$

where $i = 1,2,3$, $J = 1,2$. The coefficients $A_1$ and $A_2$ are taking the following forms respectively

$$A_1 = \sin\left(\frac{X_2}{R(T)} + \theta(T)\lambda_1(T)X_1\right) \tag{23a}$$

$$A_2 = \cos\left(\frac{X_2}{R(T)} + \theta(T)\lambda_1(T)X_1\right) \tag{23b}$$

The equilibrium configuration of the undeformed virtual SWCNT (see Fig. 4 (b)) can be established with the lattice parameters $\lambda_1(T)$, $\lambda_2(T)$, $\theta(T)$, $\xi_1^c(T)$ and $\xi_2^c(T)$ $\left(\boldsymbol{\xi}^c(T) = \left(\xi_1^c(T), \xi_2^c(T)\right)\right)$ taking their optimal values. Herein, these parameters are determined through the minimization of the surface free energy density numerically. Similarly, Arroyo and Belytschko (2004) also used the minimization process to find the inner relaxation parameters of CNTs. They pointed out that although the uniqueness of the solution for this problem could not be guaranteed, with the Newton method used, the optimal solution can be obtained within several iterations. In this study, it is worth noting that these parameters can also be obtained quietly within about 8-10 Newton iterations.

Based on the above preparation, the deformation mapping relationship from the initial planar virtual graphite sheet (see Fig. 4 (a)) to the current configuration (see Fig. 4 (c)) can be expressed as following

$$\begin{aligned} x_1 &= \lambda_1(T)X_1 + u_1 \\ x_2 &= \lambda_2(T)R(T)\sin\left(\frac{X_2}{R(T)} + \theta(T)\lambda_1(T)X_1\right) + u_2 \\ x_3 &= \lambda_2(T)R(T)\left(1 - \cos\left(\frac{X_2}{R(T)} + \theta(T)\lambda_1(T)X_1\right)\right) + u_3 \end{aligned} \tag{24}$$

where $u_1$, $u_2$ and $u_3$ are the displacement components of a point on the deformed configuration in the Euler coordinate system. The detailed expression of displacement function will be given in section 6.

## 5. Temperature-related constitutive model

The key step to obtain the equivalent temperature-related constitutive model in the quasi-continuum framework is to homogenize the free energy in the representative cell for the atomic system. With the above preparations of free energy function and the deformation mapping relationship, the constitutive model will be derived in this section.

We denote $W_{TC}$ as the surface free energy density which can be determined in the following way

$$W_{TC}\left(\boldsymbol{F}, \boldsymbol{G}, \boldsymbol{\xi}_I^c(T)\right) = \frac{U_{0I} + Q_{TI}}{\Omega_I} \tag{25}$$

where $\Omega_I$ is the area of the virtual representative cell (see Fig. 2 for reference)

$$\Omega_I = \frac{3\sqrt{3}}{4} a_0^2(T) \tag{26}$$

$U_{0I}$ denotes the interatomic potential energy over the virtual unit cell of carbon atom $I$

which can be expressed in terms of Tersoff-Brenner interatomic potential

$$U_{0I} = \frac{\sum_{J=1}^{3} V\left(\boldsymbol{r}_{IJ}^{c}(T)\right)}{2} = \frac{\sum_{J=1}^{3} V\left(\boldsymbol{F}, \boldsymbol{G}, \boldsymbol{\xi}_{I}^{c}(T)\right)}{2} \tag{27}$$

From Eq. (18), the energy $Q_{TI}$ can take the following form

$$Q_{TI} = k_B T \sum_{\beta=1}^{3} \ln\left(2 \cdot \sinh\left(\frac{\hbar\omega_{I\beta}\left(\boldsymbol{F}, \boldsymbol{G}, \boldsymbol{\xi}_{I}^{c}(T)\right)}{2k_B T}\right)\right) \tag{28}$$

From Eq. (27) and Eq. (28), the surface free energy density $W_{TC}$ can be further written as a function of parameters $\lambda_1(T)$, $\lambda_2(T)$, $\theta(T)$, $\xi_1^c(T)$ and $\xi_2^c(T)$

$$W_{TC} = W_{TC}\left(\lambda_1(T), \lambda_2(T), \theta(T), \xi_1^c(T), \xi_2^c(T)\right) \tag{29}$$

As mentioned in Section 4, from Eq. (29), these parameters can be obtained by minimizing the surface free energy density

$$\frac{\partial W_{TC}}{\partial \lambda_1(T)} = \frac{\partial W_{TC}}{\partial \lambda_2(T)} = \frac{\partial W_{TC}}{\partial \theta(T)} = \frac{\partial W_{TC}}{\partial \xi_1^c(T)} = \frac{\partial W_{TC}}{\partial \xi_2^c(T)} = 0 \tag{30}$$

Through Eq. (30), with $\hat{\boldsymbol{\xi}}_I^c(\boldsymbol{F}, \boldsymbol{G}, T)$ denoting the optimization of $\boldsymbol{\xi}_I^c(T)$, the surface free energy density $W_{TC}\left(\boldsymbol{F}, \boldsymbol{G}, \boldsymbol{\xi}_I^c(T)\right)$ can be expressed as

$$\widehat{W}_{TC}(\boldsymbol{F}, \boldsymbol{G}) = \frac{\widehat{U}_{0I} + \widehat{Q}_{TI}}{\Omega_I} = W_{TC}\left(\boldsymbol{F}, \boldsymbol{G}, \hat{\boldsymbol{\xi}}_I^c(\boldsymbol{F}, \boldsymbol{G}, T)\right) \tag{31}$$

Using Eq. (31), the first order Piola-Kirchhoff stress tensor $\widehat{\boldsymbol{P}}_T$ and the second order stress tensor $\widehat{\boldsymbol{Q}}_T$ can be derived respectively as

$$\widehat{\boldsymbol{P}}_T = \frac{\partial \widehat{W}_{TC}}{\partial \boldsymbol{F}} = \frac{1}{\Omega_I}\left(\frac{\partial \widehat{U}_{0I}}{\partial \boldsymbol{F}} + \frac{\partial \widehat{Q}_{TI}}{\partial \boldsymbol{F}}\right) \tag{32a}$$

$$\widehat{\boldsymbol{Q}}_T = \frac{\partial \widehat{W}_{TC}}{\partial \boldsymbol{G}} = \frac{1}{\Omega_I}\left(\frac{\partial \widehat{U}_{0I}}{\partial \boldsymbol{G}} + \frac{\partial \widehat{Q}_{TI}}{\partial \boldsymbol{G}}\right) \tag{32b}$$

Subsequently, from Eq. (32), we can calculate the tangent modulus tensors using the following forms

$$\widehat{\boldsymbol{M}}_{\boldsymbol{FF}} = \frac{\partial^2 \widehat{W}_{TC}}{\partial \boldsymbol{F} \otimes \partial \boldsymbol{F}} = \frac{1}{\Omega_I}\left(\frac{\partial^2 \widehat{U}_{0I}}{\partial \boldsymbol{F} \otimes \partial \boldsymbol{F}} + \frac{\partial^2 \widehat{Q}_{TI}}{\partial \boldsymbol{F} \otimes \partial \boldsymbol{F}}\right) \tag{33a}$$

$$\widehat{\boldsymbol{M}}_{FG}=\frac{\partial^2\widehat{W}_{TC}}{\partial\boldsymbol{F}\otimes\partial\boldsymbol{G}}=\frac{1}{\Omega_I}\left(\frac{\partial^2\widehat{U}_{0I}}{\partial\boldsymbol{F}\otimes\partial\boldsymbol{G}}+\frac{\partial^2\widehat{Q}_{TI}}{\partial\boldsymbol{F}\otimes\partial\boldsymbol{G}}\right) \tag{33b}$$

$$\widehat{\boldsymbol{M}}_{GF}=\frac{\partial^2\widehat{W}_{TC}}{\partial\boldsymbol{G}\otimes\partial\boldsymbol{F}}=\frac{1}{\Omega_I}\left(\frac{\partial^2\widehat{U}_{0I}}{\partial\boldsymbol{G}\otimes\partial\boldsymbol{F}}+\frac{\partial^2\widehat{Q}_{TI}}{\partial\boldsymbol{G}\otimes\partial\boldsymbol{F}}\right) \tag{33c}$$

$$\widehat{\boldsymbol{M}}_{GG}=\frac{\partial^2\widehat{W}_{TC}}{\partial\boldsymbol{G}\otimes\partial\boldsymbol{G}}=\frac{1}{\Omega_I}\left(\frac{\partial^2\widehat{U}_{0I}}{\partial\boldsymbol{G}\otimes\partial\boldsymbol{G}}+\frac{\partial^2\widehat{Q}_{TI}}{\partial\boldsymbol{G}\otimes\partial\boldsymbol{G}}\right) \tag{33d}$$

It should be pointed that the explicit expressions of $\frac{\partial\widehat{U}_{0I}}{\partial\boldsymbol{F}}$, $\frac{\partial\widehat{U}_{0I}}{\partial\boldsymbol{G}}$, $\frac{\partial^2\widehat{U}_{0I}}{\partial\boldsymbol{F}\otimes\partial\boldsymbol{F}}$, $\frac{\partial^2\widehat{U}_{0I}}{\partial\boldsymbol{F}\otimes\partial\boldsymbol{G}}\left(\frac{\partial^2\widehat{U}_{0I}}{\partial\boldsymbol{G}\otimes\partial\boldsymbol{F}}\right)$, $\frac{\partial^2\widehat{U}_{0I}}{\partial\boldsymbol{G}\otimes\partial\boldsymbol{G}}$ are very complex, and the derivation of these formulas are not given in this study. Readers can referred to Guo et al. (2008) for more details.

From the chain rule of differentiation, for the computation of $\frac{\partial\widehat{Q}_{TI}}{\partial\boldsymbol{F}}$ and $\frac{\partial\widehat{Q}_{TI}}{\partial\boldsymbol{G}}$, the key step is for computation of $\frac{\partial\widehat{\omega}_{I\beta}}{\partial\boldsymbol{F}}$ and $\frac{\partial\widehat{\omega}_{I\beta}}{\partial\boldsymbol{G}}$ ($\widehat{\omega}_{I\beta}(\boldsymbol{F},\boldsymbol{G})=\omega_{I\beta}\left(\boldsymbol{F},\boldsymbol{G},\hat{\boldsymbol{\xi}}_I^c(T)\right)$) respectively.

Through Eq. (17), the eigenequation can be expressed equivalently as follows

$$\left(\frac{1}{m_c}\frac{\partial^2U_0}{\partial\boldsymbol{x}_I\,\partial\boldsymbol{x}_I}\bigg|_{\boldsymbol{x}=\boldsymbol{x}^0}-\boldsymbol{I}_{3\times3}\left(\widehat{\omega}_{I\beta}(\boldsymbol{F},\boldsymbol{G})\right)^2\right)\boldsymbol{v}=0 \tag{34}$$

where $\boldsymbol{v}$ is an arbitrary $3\times1$ unit vector.

The first order partial derivatives of $\omega_{I\beta}$ with respect to $\boldsymbol{F}$ and $\boldsymbol{G}$ can be derived from Eq. (34) as follows

$$\frac{\partial\widehat{\omega}_{I\beta}}{\partial\boldsymbol{F}}=\frac{\frac{1}{m_c}\boldsymbol{v}^{\mathrm{T}}\frac{\partial^3U_0}{\partial\boldsymbol{x}_I\,\partial\boldsymbol{x}_I\,\partial\boldsymbol{F}}\Big|_{\boldsymbol{x}=\boldsymbol{x}^0}\boldsymbol{v}}{2\omega_{I\beta}} \tag{35a}$$

$$\frac{\partial\widehat{\omega}_{I\beta}}{\partial\boldsymbol{G}}=\frac{\frac{1}{m_c}\boldsymbol{v}^{\mathrm{T}}\frac{\partial^3U_0}{\partial\boldsymbol{x}_I\,\partial\boldsymbol{x}_I\,\partial\boldsymbol{G}}\Big|_{\boldsymbol{x}=\boldsymbol{x}^0}\boldsymbol{v}}{2\omega_{I\beta}} \tag{35b}$$

It is difficult to calculate the second order partial derivatives of $\omega_{I\beta}$ associated with $\boldsymbol{F}$ and $\boldsymbol{G}$ analytically. Herein, they are computed through the finite difference method in our numerical implementation.

## 6. Meshless numerical simulation scheme

### 6.1 Moving least-squares approximation shape function

Since the moving least-squares (MLS) approach was proposed by Lancaster and Salkauskas (1981), it has been widely used to form shape functions for meshfree methods (Belytschko et al. 1996, 1998,; Sun and Liew, 2008a, b). From the conception of MLS, the approximate nodal displacement interpolation $u$ can be expressed on the initial configuration

$$u(\boldsymbol{X},\boldsymbol{X}_0)=\sum_{i=0}^{n}p_i(\boldsymbol{X})a_i(\boldsymbol{X}_0)=\boldsymbol{p}^T(\boldsymbol{X})\boldsymbol{a}(\boldsymbol{X}_0) \tag{36}$$

where $\boldsymbol{X}=(X_1,X_2)$ is the coordinates of a material point in the reference configuration (see Fig. 4 (a)). And the coordinate space of $\boldsymbol{X}_0$ is included in the space of $\boldsymbol{X}$. The basis function $p_i(\boldsymbol{X})$ is a complete polynomial expression of order $n$ (herein $n$ is selected as 2), and $a_i(\boldsymbol{X}_0)$ is the coefficient of $p_i(\boldsymbol{X})$, as shown in follows

$$\boldsymbol{p}^T(\boldsymbol{X})=(1,X_1,X_2,X_1^2,X_1X_2,X_2^2) \tag{37a}$$

$$\boldsymbol{a}^T(\boldsymbol{X}_0)=\big(a_0(\boldsymbol{X}_0),a_1(\boldsymbol{X}_0),a_2(\boldsymbol{X}_0)\big) \tag{37b}$$

These coefficients $\boldsymbol{a}(\boldsymbol{X})$ can be obtained through a minimization process of a weighted least square fitting function $\mathcal{H}$ which taken the following form

$$\mathcal{H}=\sum_{\alpha=1}^{m}w(\boldsymbol{X}-\boldsymbol{X}_\alpha)[\boldsymbol{p}^T(\boldsymbol{X}_\alpha)\boldsymbol{a}(\boldsymbol{X})-\bar{u}_\alpha]^2 \tag{38}$$

with $m$ denoting the number of nodes in the compact support influence domain with respected to point $\boldsymbol{X}$. It should be pointed that herein we select a rectangular support as the influence domain. $\bar{u}_\alpha$ is the nodal displacement parameter associated with the $\alpha-th$ node, which is not the real nodal displacement. $w(\boldsymbol{X}-\boldsymbol{X}_\alpha)$ is the weight function as

$$w(\boldsymbol{X}-\boldsymbol{X}_\alpha) = w(d_{X_1})\cdot w(d_{X_2}) = w\left(\frac{\|X_1 - X_{\alpha 1}\|}{d_{mX_{\alpha 1}}}\right)\cdot w\left(\frac{\|X_2 - X_{\alpha 2}\|}{d_{mX_{\alpha 2}}}\right) \tag{39}$$

where $d_{mX_{\alpha 1}}$ and $d_{mX_{\alpha 2}}$ are the values of length and width of the influence domain for the $\alpha$ −th node. In this paper, we use the cubic spline function as the weight function

$$w(d) = \begin{cases} \frac{2}{3} - 4d^2 + 4d^3 & d \le \frac{1}{2} \\ \frac{4}{3} - 4d + 4d^2 - \frac{4}{3}d^3 & \frac{1}{2} < d \le 1 \\ 0 & d > 1 \end{cases} \tag{40}$$

From Eq. (38), the minimization of $\mathcal{H}$ associated with $\boldsymbol{a}(\boldsymbol{X})$ can be derived as

$$\frac{\partial \mathcal{H}}{\partial \boldsymbol{a}} = \boldsymbol{A}(\boldsymbol{X})\boldsymbol{a}(\boldsymbol{X}) - \boldsymbol{B}(\boldsymbol{X})\bar{\boldsymbol{u}} = \boldsymbol{0} \tag{41}$$

where

$$\boldsymbol{A}(\boldsymbol{X}) = \sum_{\alpha=1}^{m} w(\boldsymbol{X}-\boldsymbol{X}_\alpha)\boldsymbol{p}(\boldsymbol{X}_\alpha)\boldsymbol{p}^T(\boldsymbol{X}_\alpha) \tag{42a}$$

$$\boldsymbol{B}(\boldsymbol{X}) = [w(\boldsymbol{X}-\boldsymbol{X}_1)\boldsymbol{p}(\boldsymbol{X}_1), w(\boldsymbol{X}-\boldsymbol{X}_2)\boldsymbol{p}(\boldsymbol{X}_2), \cdots, w(\boldsymbol{X}-\boldsymbol{X}_m)\boldsymbol{p}(\boldsymbol{X}_m)] \tag{42b}$$

$$\bar{\boldsymbol{u}}^T = (\bar{u}_1, \bar{u}_2, \cdots, \bar{u}_m) \tag{42c}$$

Through Eq. (41), the coefficients $\boldsymbol{a}(\boldsymbol{X})$ can be expressed as

$$\boldsymbol{a}(\boldsymbol{X}) = \boldsymbol{A}^{-1}(\boldsymbol{X})\boldsymbol{B}(\boldsymbol{X})\bar{\boldsymbol{u}} \tag{43}$$

Substituting Eq. (43) into Eq. (36), if we denote $\boldsymbol{\phi}(\boldsymbol{X})$ as the MLS shape function, it yields that

$$u(\boldsymbol{X}) = \boldsymbol{p}^T(\boldsymbol{X})\boldsymbol{A}^{-1}(\boldsymbol{X})\boldsymbol{B}(\boldsymbol{X})\bar{\boldsymbol{u}} = \boldsymbol{\phi}(\boldsymbol{X})\bar{\boldsymbol{u}} = \phi_\alpha(\boldsymbol{X})\bar{u}_\alpha \tag{44}$$

Subsequently, the first and second order partial derivatives of $\phi_\alpha(\boldsymbol{X})$ associated with $\boldsymbol{X}$ can be calculated by

$$\phi_{\alpha,K} = \frac{\partial \phi_\alpha}{\partial X_K}, \qquad \phi_{\alpha,KL} = \frac{\partial^2 \phi_\alpha}{\partial X_K \partial X_L} \tag{45}$$

The detailed expressions of the partial derivatives for $\phi_\alpha(\boldsymbol{X})$ are not given in this study, and readers can refer to Sun and Liew (2008a) for more information.

Substituting Eq. (44) into Eq. (24), the total deformation gradient tensors $\boldsymbol{F}^{tot}$ and $\boldsymbol{G}^{tot}$ can be derived as

$$\boldsymbol{F}^{tot} = \boldsymbol{F} + \boldsymbol{F}^{def}, \qquad \boldsymbol{G}^{tot} = \boldsymbol{G} + \boldsymbol{G}^{def} \tag{46}$$

where $\boldsymbol{F}^{def}$ and $\boldsymbol{G}^{def}$ can be calculated by

$$F_{iJ}^{def} = \frac{\partial u_i}{\partial X_J} = \phi_{\alpha,J}\bar{u}_{\alpha i} \tag{47a}$$

$$G_{iJK}^{def} = \frac{\partial^2 u_i}{\partial X_J \partial X_K} = \phi_{\alpha,JK}\bar{u}_{\alpha i} \tag{47b}$$

6.2 Element Free Galerkin method

In this subsection, a temperature-dependent meshless numerical framework for simulating the buckling and post-buckling deformation behaviors of SWCNTs in the case of finite temperature is established, which is based on the proposed thermal related quasi-continuum constitutive model. The key step to construct the numerical scheme is to obtain the discrete stiffness algebraic equations from the available equilibrium equations and boundary conditions based on the theory of variational principle and numerical quadrature. Herein, the penalty function method is employed to capture the essential boundary conditions, which is proved highly effective for handling boundary problems for SWCNTs by Sun and Liew (2008a, b).

According to the principle of variation, the weak form of the governing equations can be expressed in the following way

$$\delta E = \int_{\Omega} \delta \widehat{W}_{TC} dV - \int_{\Gamma_t} t_i^P \delta u_i d\Gamma - \int_{\Gamma_t} t_i^Q N_J \frac{\partial(\delta u_i)}{\partial X_J} d\Gamma + \int_{\Gamma_u} \gamma (u_i - \tilde{u}_i) \delta u_i d\Gamma$$

$$= \int_{\Omega} \left( \left(\hat{P}_T\right)_{iJ} \phi_{\alpha,J} + \left(\hat{Q}_T\right)_{iJK} \phi_{\alpha,JK} \right) \delta \bar{u}_{\alpha i} dV - \int_{\Gamma_t} t_i^P \phi_\alpha \, \delta \bar{u}_{\alpha i} d\Gamma$$

$$- \int_{\Gamma_t} t_i^Q N_J \phi_{\alpha,J} \, \delta \bar{u}_{\alpha i} d\Gamma + \int_{\Gamma_u} \gamma (u_i - \tilde{u}_i) \phi_\alpha \, \delta \bar{u}_{\alpha i} d\Gamma$$

$$= 0 \tag{48}$$

where $i = 1,2,3,\ J,K = 1,2$. $t_i^P$ and $t_i^Q$ are the first and second order stress tractions on the surface $\Gamma_t$ associated with the $i-th$ degree of freedom respectively. $\tilde{u}_i$ denotes the boundary displacement with respect to the $i-th$ degree of freedom on the essential boundary conditions. $N_J$ is $J-th$ degree of freedom of the unit outward normal vector. $\gamma$ is a penalty factor which is selected as $10^6$ in this study.

From Eq. (48), the incremental nonlinear stiffness equations, which are solved by Newton Raphson method in this paper, can be expressed in the following compact form

$$\boldsymbol{K}(\bar{\boldsymbol{u}}_n) \cdot (\bar{\boldsymbol{u}}_{n+1} - \bar{\boldsymbol{u}}_n) = \boldsymbol{P}(\bar{\boldsymbol{u}}_n) \tag{49}$$

where $\bar{\boldsymbol{u}}_n$ and $\bar{\boldsymbol{u}}_{n+1}$ are the solutions of displacement parameters in the $n-th$ and $(n+1)-th$ iteration steps respectively. $\boldsymbol{K}(\bar{\boldsymbol{u}}_{\boldsymbol{n}})$ is the stiffness matrix which is a function of displacement parameters. $\boldsymbol{P}(\bar{\boldsymbol{u}}_n)$ is the non-equilibrium force.

From Eq. (49), the nonlinear stiffness equation can be expressed explicitly as follow

$$\left( \left( K_0(\bar{\boldsymbol{u}}_n) \right)_{\alpha i \beta j} + K_{\alpha i \beta j}^{\gamma} \right) \cdot \left( (\bar{u}_{n+1})_{\beta j} - (\bar{u}_n)_{\beta j} \right) = -\left( P_0(\bar{\boldsymbol{u}}_n) \right)_{\alpha i} - \left( K_{\alpha i \beta j}^{\gamma} \cdot (\bar{u}_n)_{\beta j} - P_{\alpha i}^{\gamma} \right) \tag{50}$$

where $\alpha$ and $\beta$ are the number of nodes in the considered system. And

$$K_{0\alpha i \beta j} = \int_{\Omega} \Big( \left(\widehat{M}_{FF}\right)_{iIjJ} \phi_{\alpha,I} \phi_{\beta,J} + \left(\widehat{M}_{FG}\right)_{iIjJK} \phi_{\alpha,I} \phi_{\beta,JK} + \left(\widehat{M}_{GF}\right)_{iIJjK} \phi_{\alpha,IJ} \phi_{\beta,K}$$

$$+ \left(\widehat{M}_{GG}\right)_{iIJjKL} \phi_{\alpha,IJ} \phi_{\beta,KL} \Big) dV \tag{51}$$

$$K^{\gamma}_{\alpha i\beta j} = \int_{\Gamma_u} \gamma\phi_\alpha\phi_\beta\delta_{ij}d\Gamma \tag{52}$$

$$P_{0\alpha i} = \int_{\Omega} \left(\left(\hat{P}_T\right)_{iI}\phi_{\alpha,I} + \left(\hat{Q}_T\right)_{iIJ}\phi_{\alpha,IJ}\right)dV - \int_{\Gamma_t} t_i^P\phi_\alpha d\Gamma - \int_{\Gamma_t} t_i^Q N_I\phi_{\alpha,I}d\Gamma \tag{53}$$

$$P^{\gamma}_{\alpha i} = \int_{\Omega} \gamma\phi_\alpha\tilde{u}_i\, d\Gamma \tag{54}$$

where $i,j = 1,2,3,\ I,J,K,L = 1,2.$ $\delta_{ij}$ denotes the Kronecker delta.

It is worth noting that the stiffness matrix $\boldsymbol{K}(\overline{\boldsymbol{u}}_{\boldsymbol{n}})$ will become non-positive definite when the bifurcation such as buckling starts to develop. In other words, the standard Newton Raphson method will be invalid and fail to find the equilibrium solution. In order to circumvent this troublesome difficulty, a positive definite stiffness matrix $\boldsymbol{K}(\overline{\boldsymbol{u}}_{\boldsymbol{n}}) + \alpha\lambda^{\boldsymbol{K}}\boldsymbol{I}$ is used to replace the non-positive $\boldsymbol{K}(\overline{\boldsymbol{u}}_{\boldsymbol{n}})$ when buckling or post buckling takes place, where $\lambda^{\boldsymbol{K}}$ is the magnitude of the most negative eigenvalue of $\boldsymbol{K}(\overline{\boldsymbol{u}}_{\boldsymbol{n}})$, $\alpha$ is a positive number to ensure stiffness matrix being positive definite and $\boldsymbol{I}$ denotes the unit matrix. After a few iterations, the non-positive $\boldsymbol{K}(\overline{\boldsymbol{u}}_{\boldsymbol{n}})$ will become positive definite, meanwhile, the modification $\boldsymbol{K}(\overline{\boldsymbol{u}}_{\boldsymbol{n}}) + \alpha\lambda^{\boldsymbol{K}}\boldsymbol{I}$ can be canceled (i.e. $\alpha = 0$). The efficiency of this simple remedy technique for investigating buckling behaviors had been verified by Liu et al. (2004) and Sun et al. (2008a, b).

The vibration frequency $\omega_{I\beta}$ is obtained through the extraction of the eigenvalue for the $3 \times 3$ vibration matrix $\frac{1}{m_c}\frac{\partial^2 U_0}{\partial x_I\, \partial x_I}$ in Eq. (17). When the buckling occurs, the vibration matrix will also be non-positive definite. We use an approximate stiffness matrix $\boldsymbol{K}^{apr}(\overline{\boldsymbol{u}}_{\boldsymbol{n}})$ which is determined in the case of without temperature effect considered (i.e. $T = 0$), so, under this circumstance, the vibration frequency will not need to calculate. A new remedy stiffness matrix $\boldsymbol{K}^{apr}(\overline{\boldsymbol{u}}_{\boldsymbol{n}}) + \alpha\lambda^{\boldsymbol{K}}\boldsymbol{I}$ can be used to replace the non-positive definite one when the buckling occurs. Although the convergence is slightly slower in the iteration step involving buckling, the equilibrium configuration and energy of the SWCNT can be obtained with higher computational efficency.

Finally, the iteration procedure of the nonlinear system is summarized as follows

(1) Firstly, assume the displacement parameter vector $\overline{\boldsymbol{u}} = \boldsymbol{0}$.

(2) If the loading steps are enforced completely, stop, otherwise continue.

(3) Calculate the stress tensors $\widehat{\boldsymbol{P}}_T$, $\widehat{\boldsymbol{Q}}_T$ and the tangent modulus tensors $\widehat{\boldsymbol{M}}_{\boldsymbol{FF}}$, $\widehat{\boldsymbol{M}}_{\boldsymbol{FG}}$, $\widehat{\boldsymbol{M}}_{\boldsymbol{GF}}$, $\widehat{\boldsymbol{M}}_{\boldsymbol{GG}}$.

(4) Calculate the shape function $\boldsymbol{\phi}$ and its partial derivatives $\boldsymbol{\phi}_{,I}$ and $\boldsymbol{\phi}_{,IJ}$.

(5) Calculate the natural boundary condition.

  (a) If the natural boundary condition is enforced in the first iteration step of one loading step, $\boldsymbol{t}^P = \boldsymbol{0}$ and $\boldsymbol{t}^Q = \boldsymbol{0}$.

  (b) Form the force vector $\boldsymbol{P}_0$.

(6) Enforce the essential boundary condition.

  (a) If the essential boundary condition is enforced in the first iteration step of one loading step, $\widetilde{\boldsymbol{u}} = \boldsymbol{0}$.

  (b) Calculate the stiffness matrix $\boldsymbol{K}^{\gamma}$.

  (c) Assemble the non-equilibrium force vector $\boldsymbol{P}(\overline{\boldsymbol{u}}_n)$.

(7) Assemble the total stiffness matrix $\boldsymbol{K}(\overline{\boldsymbol{u}}_{\boldsymbol{n}})$.

If $\boldsymbol{K}(\overline{\boldsymbol{u}}_{\boldsymbol{n}})$ is positive definite, continue.

If $\boldsymbol{K}(\overline{\boldsymbol{u}}_{\boldsymbol{n}})$ is nonpositive definite, $\boldsymbol{K}(\overline{\boldsymbol{u}}_{\boldsymbol{n}})$ is replaced by $\boldsymbol{K}^{apr}(\overline{\boldsymbol{u}}_{\boldsymbol{n}}) + \alpha\lambda^{\boldsymbol{K}}\boldsymbol{I}$, continue.

(8) Solve the equation $\boldsymbol{K}(\overline{\boldsymbol{u}}_n) \cdot (\overline{\boldsymbol{u}}_{n+1} - \overline{\boldsymbol{u}}_n) = \boldsymbol{P}(\overline{\boldsymbol{u}}_n)$.

(9) $\overline{\boldsymbol{u}}_n = \overline{\boldsymbol{u}}_{n+1}$.

(10) Judge whether the convergence criterion is satisfied.

Yes, directly go to (11).

No, go to (2).

(11) Calculate the displacement vector $\boldsymbol{u} = \boldsymbol{\phi}(\boldsymbol{X})\overline{\boldsymbol{u}}$, and enforce the next loading step, go to (2).

## 7. Numerical simulation

In this section, with the proposed thermal-related meshless computational scheme based on the temperature dependent quasi-continuum constitutive model, the buckling and postbuckling deformation behaviors of SWCNTs subjected to axial compression and torsion deformations at finite temperature are investigated. And the influence of tube radii on the critical strains of zigzag and armchair SWCNTs are also studied. These problems are treated as quasi-static and isothermal cases (the ambient temperature is not very high). The equilibrium solution is calculated in each loading step.

### 7.1 Compressed (10, 0) SWCNTs

Firstly, for the purpose of verifying the validity of the proposed computational model, the buckling deformation behaviors of a (10, 0) SWCNT under axial compression at 300K (i.e. at room temperature) are investigated. The effective length of the nanotube is 5.3 nm (a total of 420 atoms). One end of the tube is fixed, and the axial compression loading is enforced on the free end incrementally. At the early stage, one loading step is chosen as 0.1nm, and then when the deformation of the SWCNT closes to buckling, 0.01nm per loading step is used. The symbol $n \times m$ denotes the number of meshless nodes distribute uniformly along the longitudinal and lateral direction of initial virtual planar graphene sheet respectively.

Fig. 5 (a) shows the variations of strain energy density of the (10, 0) SWCNT at 300K under different axial compression strain. It can be observed that the strain energy per atom increases quadratically with the axial compression strain increasing and then when the axial strain reaches a critical value, the accumulated deformation energy will release partly. Beyond this critical strain, the strain energy density increases linearly. For comparison, the simulation results obtained through the molecular dynamic (MD) method reported by Zhang et al. (2009) are also depicted in Fig. 5 (a). Before buckling occurs, although the results obtained by our proposed numerical model are slightly smaller than the simulation values provided by Zhang et al. (2009), the trend of the variations of the strain energy density and the value of critical strain are consistent with each other. However, the discrepancy between the results obtained by our model and by the MD method is large after buckling. The possible source for the bifurcation might be that our constitutive model is more rigid than the atomistic method. It is because that if atom $I$ is considered to calculate the free energy in the representative cell, the other atoms are fixed (see Fig. 2 for reference).

The effect of the number of the selected meshless nodes on the buckling deformation behavior of the SWCNT is also shown in Fig. 5 (a). It can be observed that as buckling taking place, the amount of the energy jump will increase with the number of meshless nodes increasing. It is reasonable from the point of numerical computation that the structure will be more compliant with the number of nodes increasing. The critical strain changes from 6.6% to 6.8% with the number of nodes increasing from $16 \times 12$ to $31 \times 20$. It is interesting to note that although the number of nodes has somewhat influence on the critical strain, these critical values match perfectly with that of 6.714% reported by Zhang et al. (2009).

The corresponding buckling deformation patterns of the (10, 0) SWCNT for different number of nodes are shown in Fig. 5 (c), (d) and (e). Obviously, with the nodes increasing from $16 \times 12$ to $31 \times 20$, the local deformations of the SWCNT will be more severely. The deformation patterns as shown in Fig. 5 (c) and Fig. 5 (d) are very similar to the MD simulation result (see Fig. 5 (b) for reference). Accordingly, with fewer number of meshless nodes used, a reasonable simulation result can be obtained by the proposed thermal-related numerical framework.

7.2 Compressed (7, 7) SWCNT

In order to further verify the efficiency of the numerical framework, the buckling and postbucking behaviors of a SWCNT undergoing severe deformation will be tested in this subsection. A (7, 7) armchair SWCNT 6.3nm long is selected for axial compression test. The number of the meshless nodes are $37 \times 25$. One end of the SWCNT is fixed, the load imposed on the free end in the same way as shown in subsection 7.1. The variations of strain energy per atom of the SWCNT associated with axial compression strain at different temperature (100K, 300K and 500K) are shown in Fig. 6 (a). At 100K, the first buckling critical strain is 6.023%, and the postbuckling critical strains are 8.876% and 9.668%. At 300K, the critical strains are 5.984%, 8.819% and 9.606%. And at 500K, they become 5.891%, 8.527% and 9.457% respectively. It is obviously that, with temperature increasing from 100K to 500K, the value of critical strain will decrease. Zhang and Shen (2006) and Zhang et al. (2009) are also obtained a similar trend of the variation of critical strain for SWCNTs with MD method employed.

Fig. 6 (b) shows the buckling and postbuckling deformation patterns of the SWCNT corresponding to the critical strain points which are depicted in Fig. 6 (a). For this (7, 7) SWCNT with a relatively small aspect ratio, the shell-like buckling patterns are observed. These buckling modes are very similar to those reported by Yakobson et al. (1996) with MD method used (see Fig. 6 (c) for an illustration). It is obvious that, from the buckling pattern (1) to (3), the local buckling occurs firstly, and then with the deformation increasing, the SWCNT will fall into global buckling status.

7.3 Critical strains of SWCNTs under compression

In this subsection, the influence of nanotube radius on the critical strain of two types SWCNTs ((n, n)armchair and (n, 0) zigzag SWCNTs) under axial compression at 0K and 300K are studied. The effective length of all the nanotubes considered in this test is 10.1nm. Fig. 7(a) shows the variations of critical strain as functions of tube radius for SWCNTs with different chirality. It can be seen that when the nanotube radius is smaller than 0.56nm, the critical strain increases linearly with the tube radius increasing and then a tip is reached. However, the trend of its variation is opposite and the critical strain decreases very quickly with the tube radius continuously increasing. For the tube radius is larger than 0.88nm, the critical strain decreases in approximate index form and then a saturated value appears when the radius is larger than 2nm. It is obviously that the critical strain is not sensitive to the chirality of SWCNTs. For the comparison, the results obtained through MD method by Cornwell and Wille (1998) at 0.005K (very close to absolute zero temperature) are also depicted in Fig. 7 (a). Our results are in good agreement with those obtained by MD method. Fig. 7 (a) also shows that the critical strains predicted at 300K are smaller than the corresponding ones at 0K. It indicates that the SWCNTs will be softening with the temperature raising.

Three buckling modes of SWCNTs with different aspect ratio are shown in Fig. 7 (b), (c) and (d). If the radii of SWCNTs range from 0.3nm to 0.56nm, the SWCNTs will undergo beam-like buckling modes which are very similar with those given by Wang et al. (2007) and Wang et al. (2010) (see Fig. 7 (b) for reference). If a SWNCT falls into the range of the tube radius between 0.56nm and 0.88nm, the shell-like buckling mode can be observed (as shown in Fig. 7 (c)). In addition, when the radius of a SWCNT is larger than 0.88nm the can-like buckling deformation pattern will onset (see Fig. 7 (d)).

It is interesting to note that the variations of the critical strain can be divided into three stages such as b, c and d (see Fig. 7 (a) for an illustration). In the stage b (i.e. the radii of SWCNTs in the range between 0.3nm to 0.56nm), the buckling patterns of SWCNTs are like beams (see Fig. 7 (b) for reference) which are very similar with those given by Wang et al. (2007) and Wang et al. (2010). If a SWNCT falls into the range of the tube radius between 0.56nm to 0.88nm (in stage c) a shell-like buckling mode will appear at the critical strain point (as shown in Fig. 7 (c)). In addition, when the radius of a SWCNT is larger than 0.88nm (i.e. in stage d) a can-like buckling deformation will onset (as shown in Fig. 7 (d)). The transformation of bucking modes for SWCNTs with different radii predicted by our numerical model is very similar to that reported by Cornwell and Wille (1998).

7.4 Twisted (21, 0) SWCNT

In this test, a (21, 0) zigzag SWCNT 11.2nm long (a total of 1890 atoms) is chosen for twisting test at room temperature (300K). The rotation is applied on the two ends of the SWCNT in opposite direction with respect to the axis of the undeformed nanotube. And the loading is enforced quasi-statically with one degree per step. For comparison, $37 \times 30$ and $41 \times 21$ meshfree nodes are used to investigate the influence of the number of nodes on deformation behaviors of the considered SWCNT. Fig. 8 (a) shows the variations of strain energy density with respected to twisting angle at 300K. It can be seen that in the small strain range (i.e. twisting angle per length less than $2^{\circ}/\text{nm}$), the strain energy density increases quadratically with the twisting angle increasing. However, when the twisting angle per length is larger than the critical value of $2^{\circ}/\text{nm}$, the strain energy density increases nearly linearly. For the comparison, the MD simulation results reported by Zhang et al. (2006) are also depicted in Fig. 8 (a). The trend of the variations of strain energy density predicted by our model is in good agreement with that obtained through MD method. The energy difference between the two methods becomes large with an increasing of twisting angle. As aforementioned, the most probably source is that our quasi-continuum model is more rigid than the MD model.

As shown in Fig. 8 (a), it is interesting to note that the strain energy-strain curve is extremely smooth unlike the case of axial compression, even though the buckling occurs. This matches well with those reported by Zhang et al. (2009) and Khademolhosseini et al. (2010). The possible reason for this phenomenon is that the increased and released deformation energy of the SWCNT under torsion keeps equilibrium. And a reasonable result for the SWCNT under torsion can be obtained with fewer meshless nodes used. Due to the absence of non-bonded interactions, the buckling pattern of the (21, 0) SWCNT at twisting anlge per length of $3.98°/\text{nm}$ predicted by our model is slightly slimmer than the MD simulation result (see Fig. 8 (b) and (c)).

## 8. Conclusion

In the present paper, we developed a thermal-related meshless numerical scheme based on the thermo-related quasi-continuum theory to simulate the large deformation behaviors of armchair and zigzag SWCNTs at finite temperature. Based on Helmholtz free energy and THCB rule employed, the thermal effect and the curvature effect can be taken into consideration conveniently and highly cost-effective. The mesh-free method is a reasonable candidate for the implementation of the quasi-continuum model with requirement of higher order continuity. The axial compression and torsion tests for two types SWCNTs implement numerically with the proposed numerical framework at finite temperature. Compared with the molecular dynamic modeling, good simulation results can be predicted with fewer meshless nodes used in our numerical simulation. It is foreseeable that if the non-bonded interactions are considered in our quasi-continuum model, the simulations for large deformation behaviors of SWCNTs will be more perfect.

It is worth noting that this developed numerical scheme is not restricted to CNTs. The thermo-mechanical properties of graphene can also be simulated by this numerical framework, and it will be presented in a further work. In addition, it is because the bond rearrangements or defects cannot be conducted by this quasi-continuum model, a hybrid numerical scheme combined with atomistic model can be developed in the days ahead.

**Acknowledgements**

The financial supports from the National Natural Science Foundation (10925209，10772037, 10472022, 10721062) and the National Key Basic Research Special Foundation of China (2006CB601205) are gratefully acknowledged.

**Figure Captions**

Fig. 1.

Deformation Mapping of standard Cauchy-Born rule

Fig. 2.

(a) representative cell corresponding to a carbon atom $I$, (b) schematic diagram of inner displacement $\boldsymbol{\eta}$ vector.

Fig. 3.

The vibration centers of initial equilibrium graphene and the virtual bond length $a_0$ at temperature $T$.

Fig. 4.

Deformation mapping from virtual graphene sheet to current configuration: (a) an initial undeformed virtual graphene; (b) the undeformed SWCNT rolled by (a) with $\phi$ denoting the twisting angle; (c) the deformed SWCNT under certain boundary condition.

Fig. 5.

Comparison between meshless simulation and molecular dynamic simulation of (10, 0) SWCNT at 300K, and the influence of the number of meshless nodes on buckling patterns, (a) strain energy per atom versus axial compression strain, (b) buckling pattern obtained by Zhang et al. (2009), (c) buckling pattern of $16 \times 12$ nodes, (d) buckling pattern of $21 \times 15$ nodes, (e) buckling pattern of $31 \times 20$ nodes.

Fig. 6.

Buckling and postbuckling behaviors of the (7, 7) SWCNT under axial compression, (a) the variations of strain energy per atom with axial compression strain at different absolute temperature, (b) the buckling and postbuckling patterns, (c) the buckling and postbuckling patterns obtained by Yakobson et al. (1996).

Fig. 7.

Critical strain of two type SWCNTs, (a) relationship between critical strain and tube radius at 0K and 300K, (b), (c) and (d) are buckling patterns with different radius.

Fig. 8.

Twisted the (21, 0) SWCNT at 300 K: (a) variations of strain energy with twisting angle unit length; (b) and (c) are buckling patterns at $3.98°/\text{nm}$ obtained by our model and the molecular dynamic model (Zhang et al. 2006) respectively.

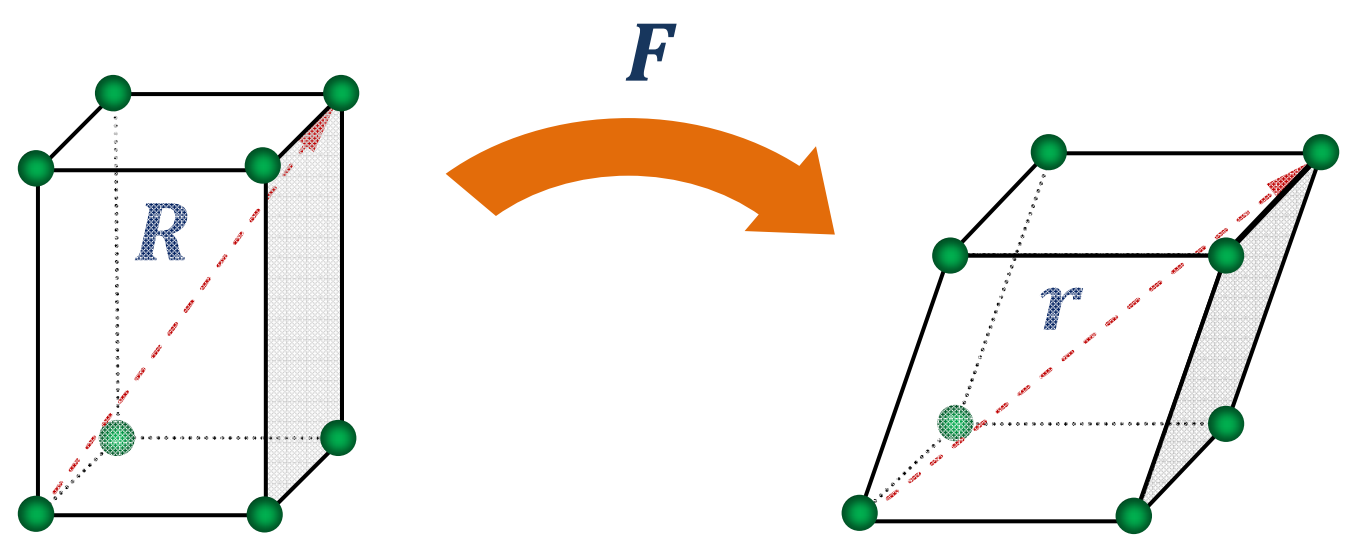

**Fig. 1.**

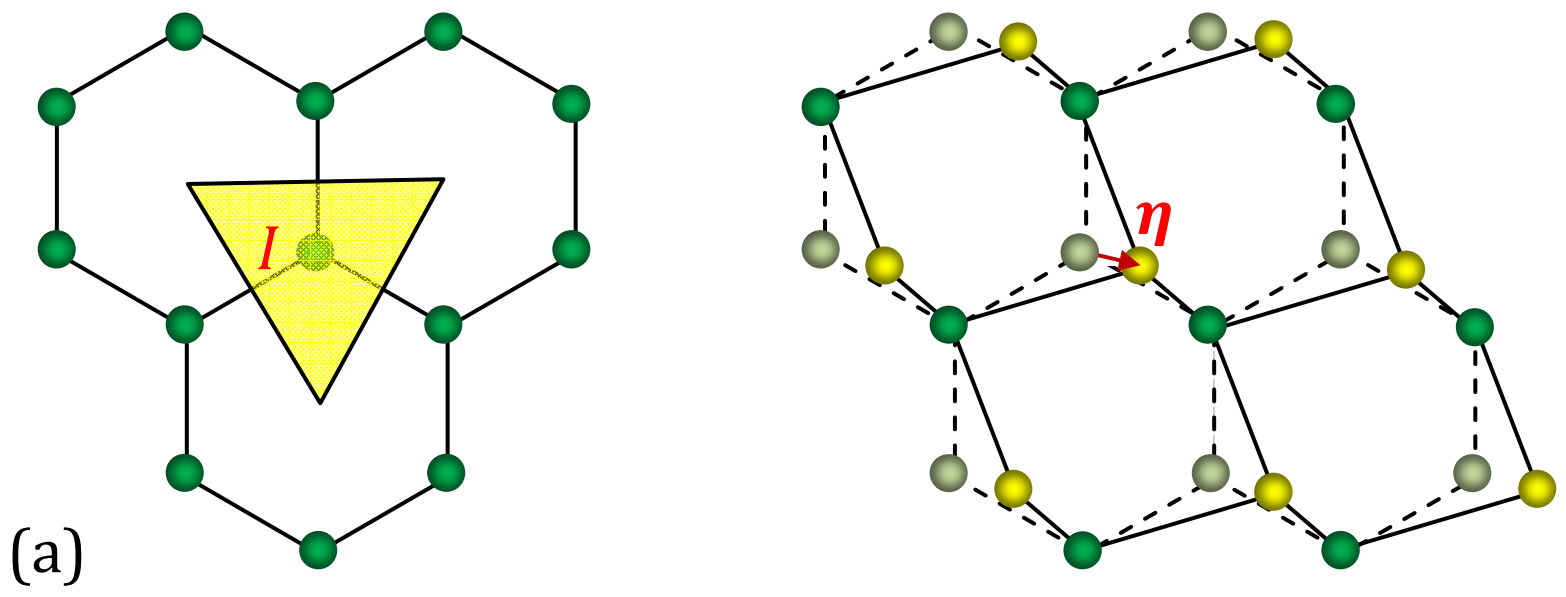

**Fig. 2.**

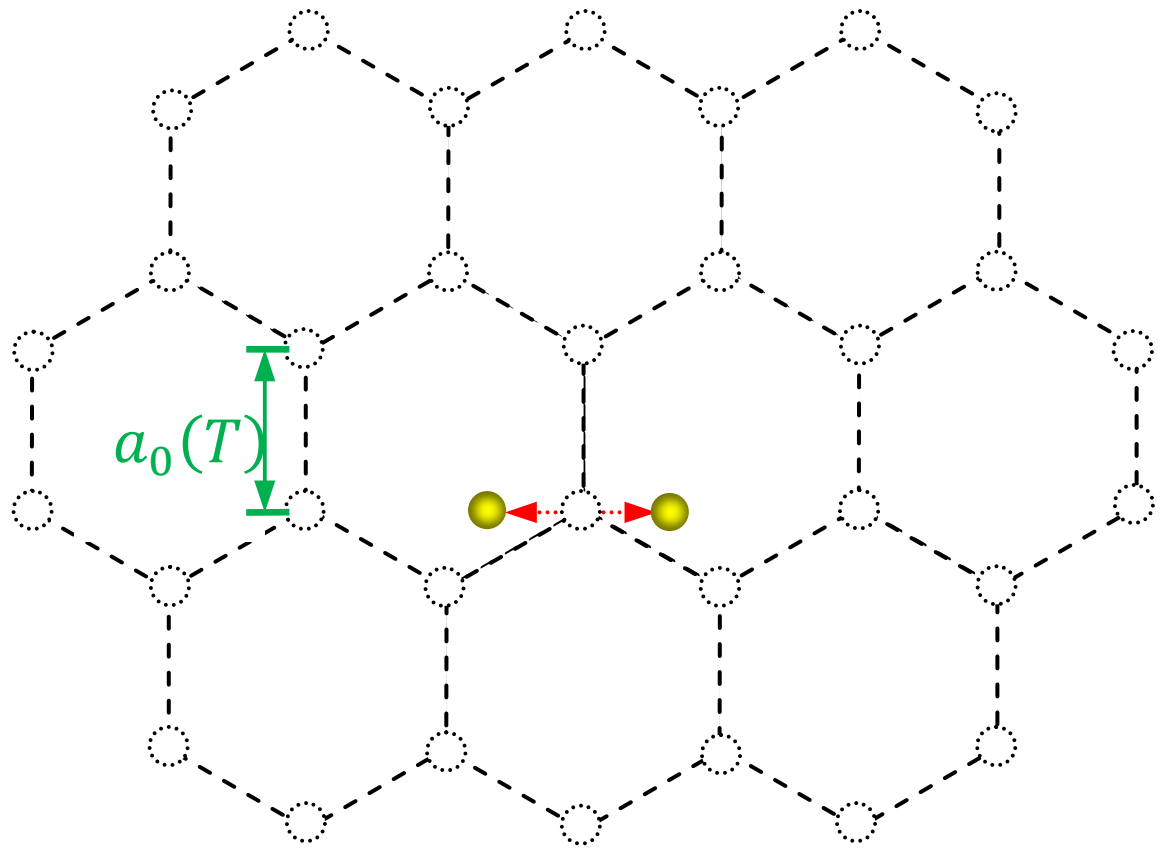

**Fig. 3.**

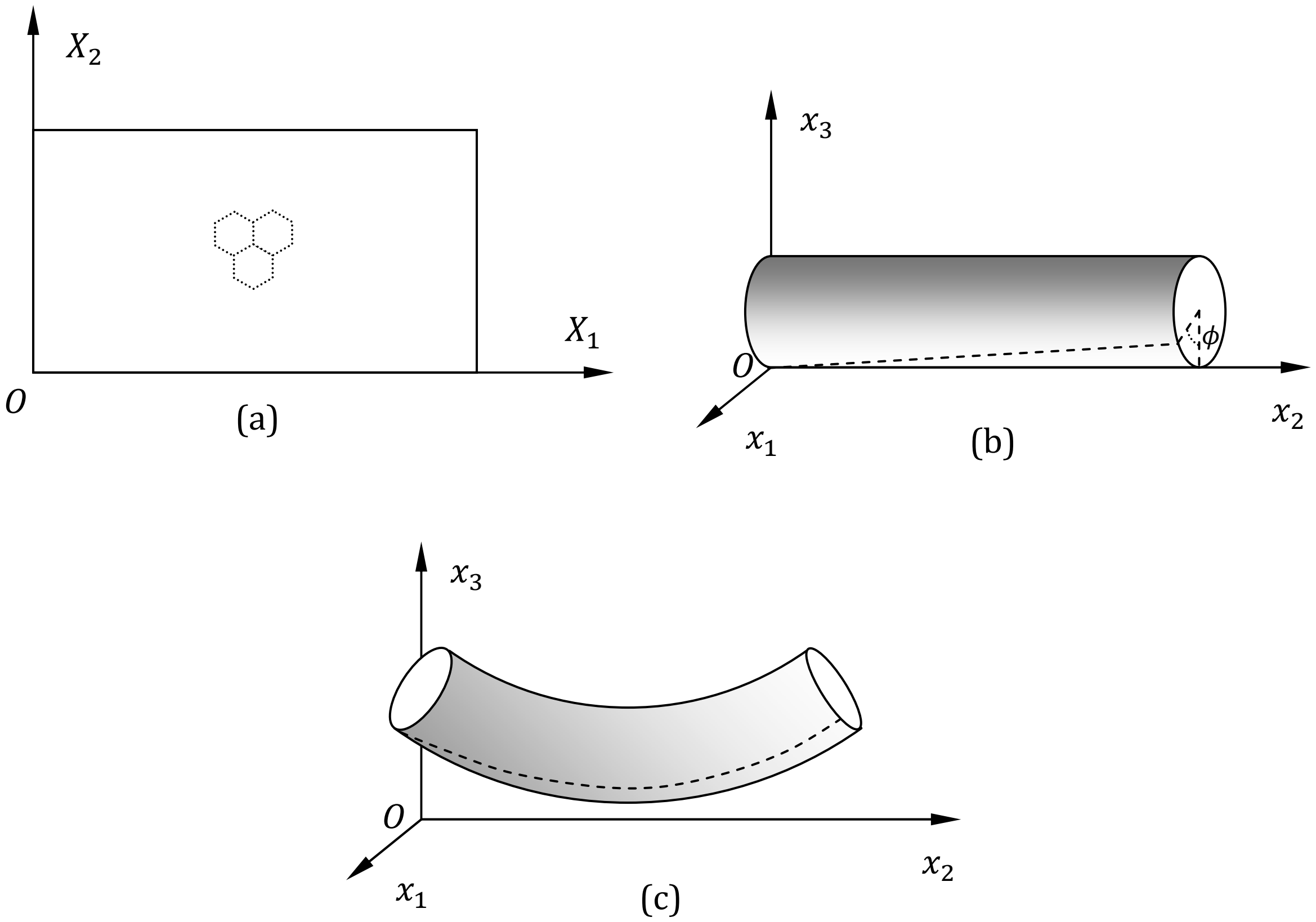

**Fig. 4.**

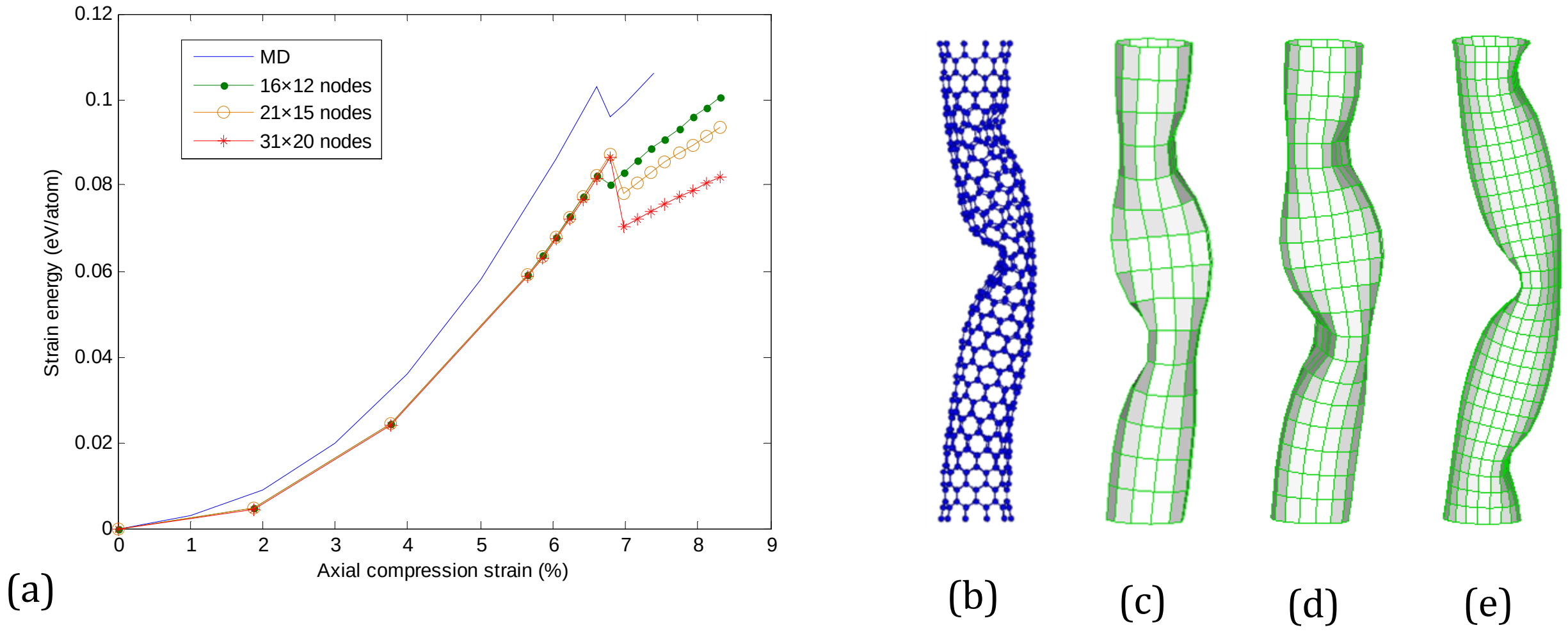

MD
16×12 nodes
21×15 nodes
31×20 nodes
Strain energy (eV/atom)
Axial compression strain (%)
0.12
0.1
0.08
0.06
0.04
0.02
0
0 1 2 3 4 5 6 7 8 9
(a)
(b)
(c)
(d)
(e)

**Fig. 5.**

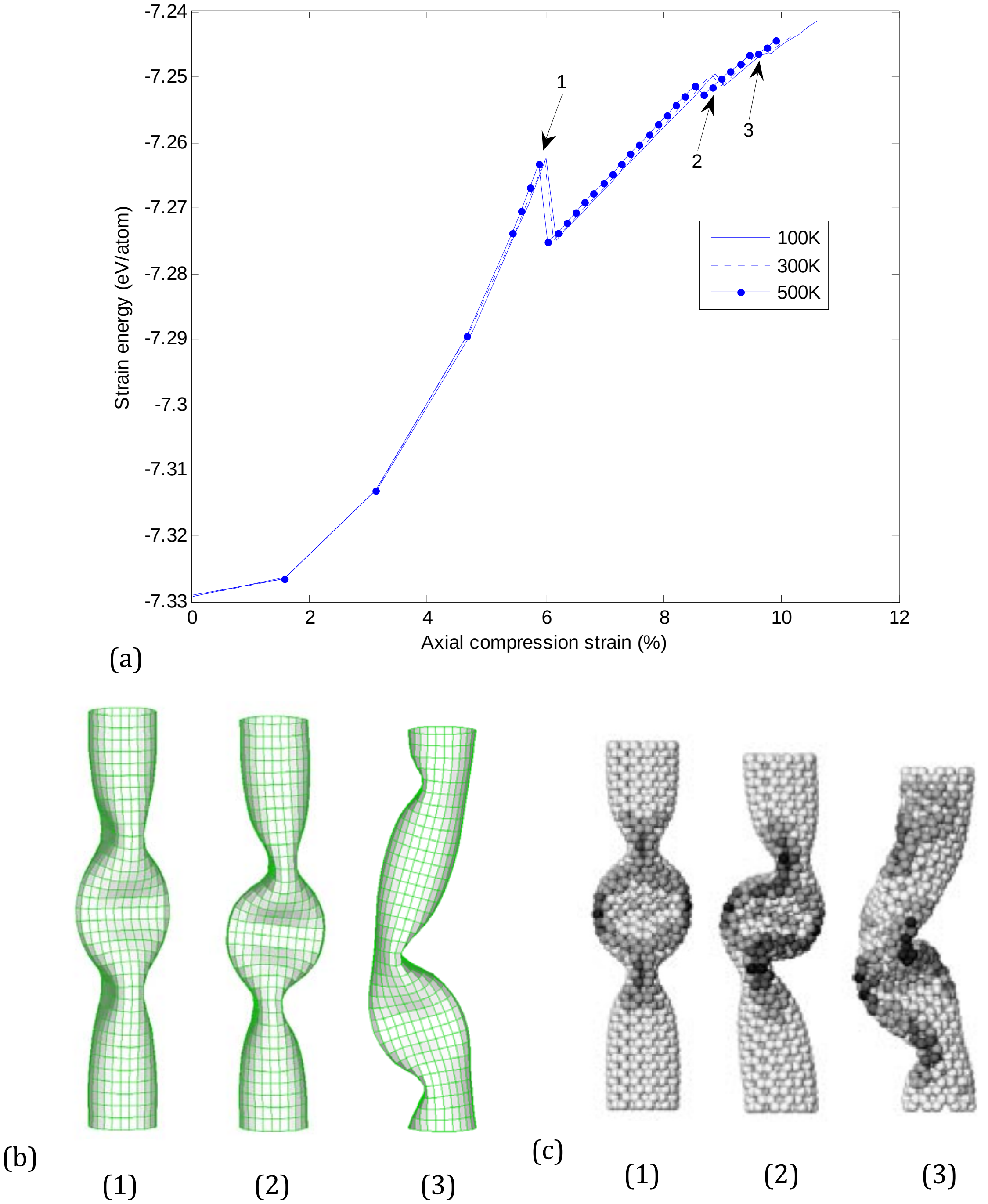

**Fig. 6.**

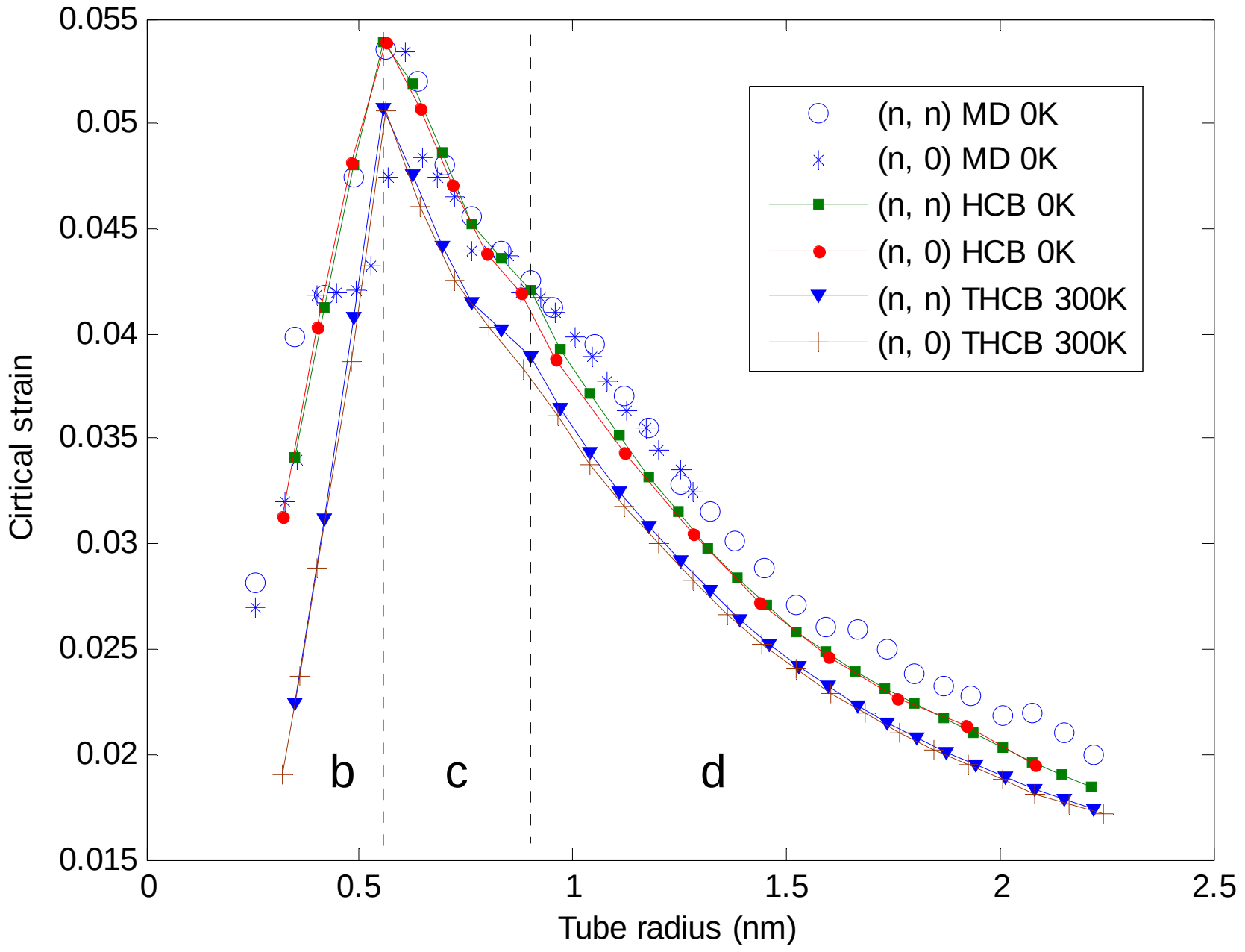
0.055
0.05
0.045
0.04
0.035
0.03
0.025
0.02
0.015
Cirtical strain
0
0.5
1
1.5
2
2.5
Tube radius (nm)
(n, n) MD 0K
(n, 0) MD 0K
(n, n) HCB 0K
(n, 0) HCB 0K
(n, n) THCB 300K
(n, 0) THCB 300K
b
c
d

(a)

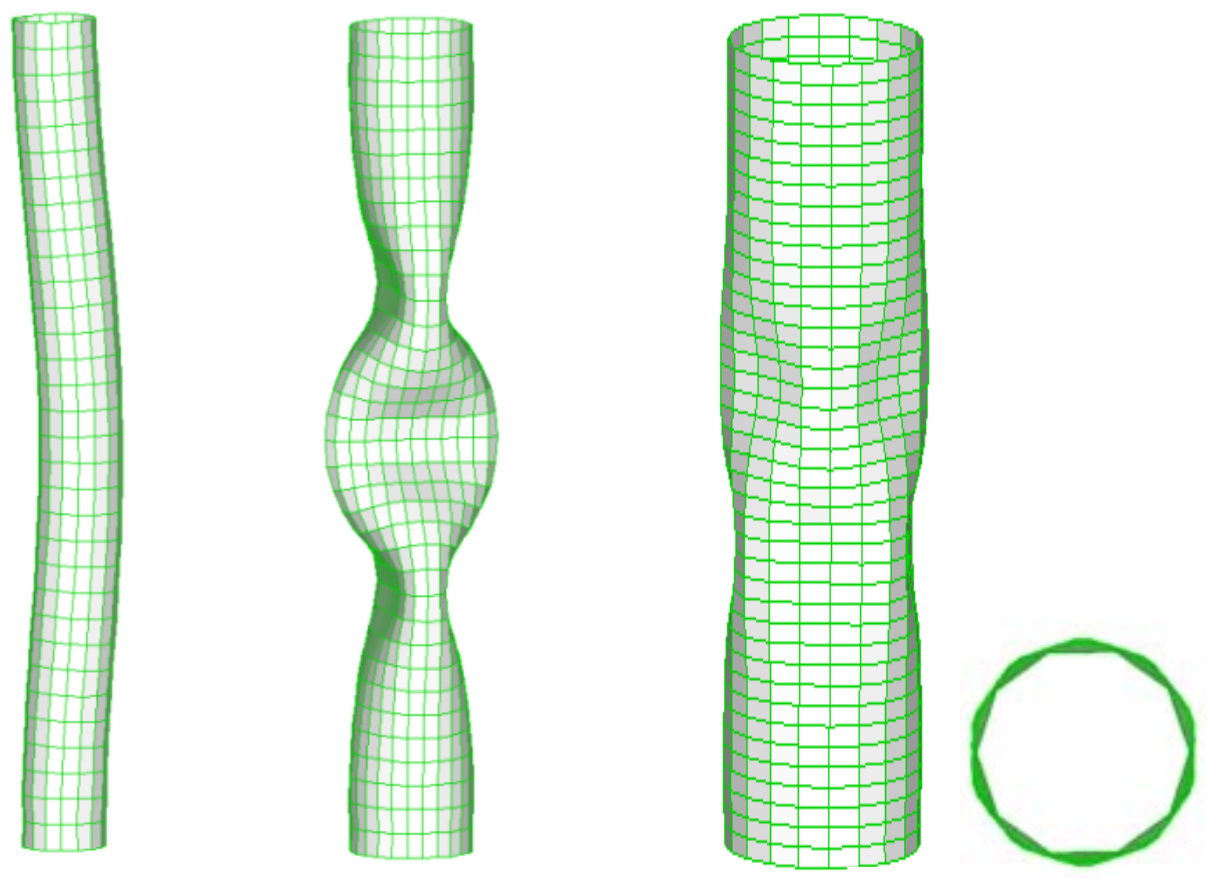

(b) (c) (d)

**Fig. 7.**

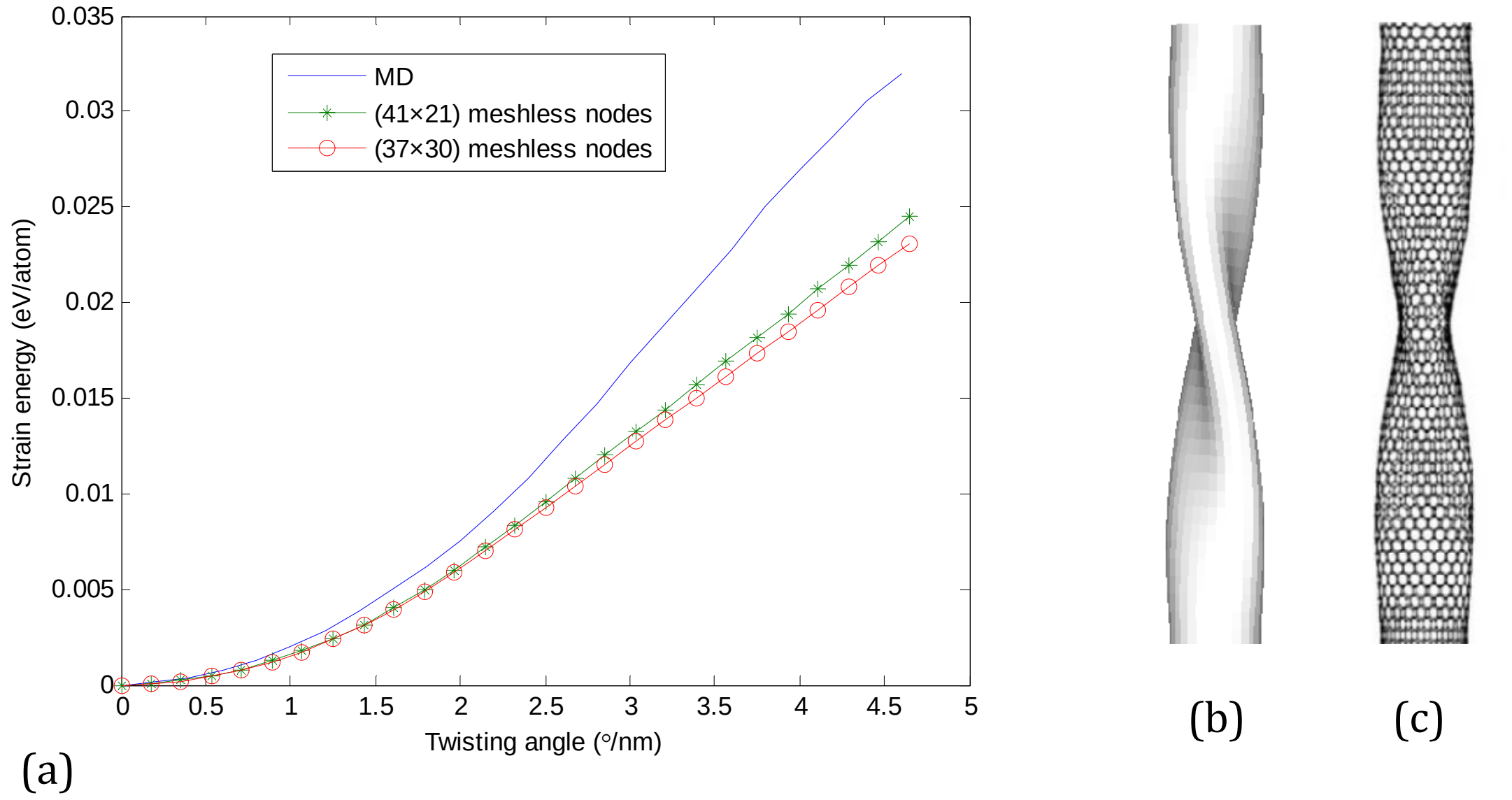

**Fig. 8.**